# One-photon wavepacket interacting with two separated atoms in a one-dimensional waveguide: Influence of virtual photons.

S. Derouault and M. A. Bouchene*

*Laboratoire de Collisions Agrégats Réactivité, C.N.R.S. UMR 5589, IRSAMC,
Université Paul Sabatier, 118 Route de Narbonne, 31062 Toulouse Cedex 4, France*

*E-mail address: aziz@irsamc.ups-tlse.fr

*Tel : (33) 5 61 55 60 02 ; fax : (33) 5 61 55 83 17

## Abstract

We present a theoretical study of one photon wave-packet scattered by two atoms in one dimensional waveguide. We investigate the role of non rotating wave approximation terms to take into account correctly for the effects of the virtual photons that are exchanged between the atoms. These terms are shown to influence drastically the reflected and the transmitted fields, imposing strict constraints on their temporal envelopes.

# I- **Introduction:**

The control of the interaction between light and matter is a research area undergoing continuous evolution because of the appearance of ever new challenges. A recent issue is the realization of all-optical quantum devices in one-dimensional waveguide for quantum information purposes [1-3]. Recent experimental progress in the designing of these systems [4-17] and the possibility to reach the strong interaction regime between photons and atoms (or artificial atoms) open new perspectives, allowing the controllable transport of the flying qubits (photon) and the realization of fundamental quantum information operations [1-3, 18-22]. Beside these challenging, the interaction of light and a collection of atoms in such systems represents by its own an interesting new theoretical problem. The photon scattering by a single atom in a 1-d waveguide has been studied by Domokos et al. in a two-level system [23] using a Heisenberg approach whereas spectral studies involving different experimental configurations have also been realized in [2, 18], and the case for three-level atoms has been studied by Witthaut and Sorensen [24]. The extension of these studies to systems with two artificial atoms and an array of N artificial atoms have been also investigated [25-30]. However, all these studies were restricted to the regime where the rotating wave approximation (RWA) is done. Introduction of frequencies cut-off, extension to negative frequencies are some procedures that are generally invoked to justify the neglect of far resonance frequencies or to recover finite coupling. In the case of two atoms in a dispersionless waveguide, non-RWA contributions cannot be neglected and are essential for a correct treatment of the problem and a deep understanding of virtual photon effects on the system dynamics. These features are well known in the field of superadiance since the exhaustive works of Friedberg, Hartmann and Manassah [31, 32], Milloni and Knight [33] and others [34, 35] following the pioneering work of Dicke [36]. The influence of virtual photons on the collective spontaneous emission of a photon wave packet by a (3-d) cloud of dense atoms has recently received a great deal of attention [37-39]. This problem is particularly rich in new striking quantum effects like collective lamb-shift, collective encoding, entanglement and directive photon reemission [40-43]. This interaction is associated with atomic shifts that modify the dynamics even for large samples.

Here, we present a detailed study of the scattering of a one photon wave packet by a system of two atoms in a lossless and dispersionless 1-d waveguide, taking into account the effects of the virtual photons. RWA is not done and we are interested in both the temporal and the spectral behavior of the scattered field. We show that both the atomic and the field dynamics dependent strongly on the nature (real or virtual) of photons exchanged by the atoms and we clarify the role of each. Moreover, we establish the expression of the effective coupling between atoms and we show that it results from a subtle interference effect between parts of virtual photons. We discuss the consequences of using RWA and introducing artificial frequencies cut-off. We demonstrate the important result that the central wave packet frequency is always reflected only if non-RWA terms are taken into account. An additional feature in our approach is the development of a "time-dependent" point of view for the interaction. We show that the total reflection of the resonant frequency is related to the specific behavior of the temporal envelopes of the reflected and transmitted fields. Moreover, the transmitted wavepacket obeys a strong constraint that forces the electric field to distort so that its pulse- area (e.g. integral of the electric field envelope) vanishes, whereas for the reflected field the pulse area is opposite to the incident one. This feature was already pointed in our previous study of photon scattering by a *single* atom in 1-d waveguide [44] and turn to be fruitful to understand straightforwardly some temporal shaping effects. The time dependent approach is only little addressed in quantum optics, in contrast with semiclassical optics, where intensive studies have been carried out leading to fascinating experiments for optical control and manipulation of quantum systems [45].

## II- **The theoretical model :**

We consider two identical atoms that interacts *resonantly* with a one-photon wave-packet propagating in the +z direction of an infinite lossless waveguide (figure 1). The transverse dimension of the waveguide $d$ is assumed to be much smaller than $\lambda_0$ (the resonant wavelength) and the interatomic distance $l$ (e.g. $d << \lambda_0, l$). An important consequence is that the electrostatic dipole-dipole interaction between the atoms is strongly inhibited in the waveguide and will be neglected through the paper [46, 47]. Moreover, the atoms no longer radiate outside the z direction and the field remains uniform in the longitudinal direction of propagation [23, 24]. The confinement of light in this waveguide ensures also that the strong interaction regime between the atoms and the photons can be realized.

The identical atoms are labelled $j = 1, 2$ and are each modelized by a two-level system (ground states $|a_j\rangle$ and excited states $|b_j\rangle$ with eigenfrequencies 0 and $\omega_0$ respectively). In our formalism, the Hamiltonian of the system $\hat{H}$ can be separated into three terms $\hat{H} = \hat{H}_{atomic} + \hat{H}_{field} + \hat{H}_{inter}$. In this notation, $\hat{H}_{atomic} = \sum_{j=1}^{2} \hbar\omega_0 |b_j\rangle\langle b_j|$ is the Hamiltonian of the free atoms, $\hat{H}_{field} = \int_{-\infty}^{+\infty} (\hbar\omega_k) \hat{a}_{k_z}^{\dagger} \hat{a}_{k_z} dk_z$ is the Hamiltonian of the free field with $\omega_k = c|k_z|$ and $\hat{a}_{k_z}$ the photon annihilation operator that follows the usual bosonic commutation rules $[\hat{a}_{k_z}, \hat{a}_{k_z'}^{\dagger}] = \delta(k_z - k_z')$. $\hat{H}_{inter} = \sum_{j=1}^{2} \int_{-\infty}^{+\infty} (\hbar g_k)(\hat{a}_{k_z}^{\dagger} e^{-ik_z z_j} + \hat{a}_{k_z} e^{ik_z z_j})(\hat{\sigma}_j + \hat{\sigma}_j^{\dagger}) dk_z$ is the interaction Hamiltonian written in the Coulomb jauge with $z_j$ the position of atom $j$ (with $z_2 - z_1 = l$), $g_k = \dfrac{(\omega_0 d_{ab})}{(4\pi\varepsilon_0 (\hbar\omega_k) A)^{1/2}}$ the coupling constant ($A$ is the effective transverse guide section, $d_{ab}$ the dipole moment) and $\hat{\sigma}_j = |a_j\rangle\langle b_j|$ the lowering operator. Note that since the coupling $g_k$ diverges in the infrared and decreases only slowly in the UV domain, we cannot neglect the contributions of any frequency and RWA cannot be done [31-35, 37-42, 48].

With initially the atoms in the ground state, and for the second order in the interaction Hamiltonian, the wavefunction $|\psi\rangle(t)$ of the whole system (atoms+field) can be formally expanded as :

$$|\psi\rangle(t) = \int_{-\infty}^{+\infty} \alpha_{k_z}(t) e^{-i\omega_k t} |a_1, a_2, 1_{k_z}\rangle dk_z + \sum_{j=1}^{2} \beta_j(t) e^{-i\omega_0 t} |a_{j'\neq j}, b_j, 0\rangle + \int_{-\infty}^{+\infty} \gamma_{k_z}(t) e^{-i(2\omega_0 + \omega_k)t} |b_1, b_2, 1_{k_z}\rangle$$
$$+ \sum_{j=1}^{2} \int_{-\infty}^{+\infty} dk_z \int_{-\infty}^{+\infty} dk_z' \eta_{j,k_z,k_z'}(t) e^{-i(\omega_k + \omega_{k'} + \omega_0)t} |b_j, 1_{k_z}, 1_{k_z'}\rangle \quad (1)$$

The two first terms corresponds to states with an excitation number equal to one. In the first term, we have states with one photon in the field and both atoms in the ground level whereas in the second term, we have states with only one atom (*j*) in the excited state and no photons in the field. The last two terms correspond to an excitation number of three. The third term describes the situation where both atoms are excited and there is one photon in the field, whereas the last term corresponds to the situation with one excited atom (*j*) and two photons in the field. These states are necessary for the correct treatment of virtual photon and the collective Lamb-shift effects [37-38, 42].

The evolution of the system is determined by the Schrödinger equation $i\hbar\dfrac{d|\psi\rangle}{dt} = \hat{H}|\psi\rangle$ with the initial conditions $\beta_j(t \to -\infty) = \eta_{j,k_z,k_z'}(t \to -\infty) = \gamma_{j,j',k_z}(t \to -\infty) = 0$ and

$\alpha_{k_z}(t \to -\infty) = \sqrt{\dfrac{c}{\Delta}}\sqrt{\dfrac{1}{2\pi}} e^{-(\omega_k - \omega_0/\Delta)^2}$ ($\Delta$ is the spectrum bandwidth). Using equation (1), we obtain the following set of equations for the amplitudes:

$$i\dot{\alpha}_{k_z}(t) = \sum_{j=1,2}\left[ g_k\left(\beta_j(t)e^{-i(\omega_0-\omega_k)t} e^{-ik_z z_j}\right) + \left(2\int_{-\infty}^{+\infty} g_{k'}\, \eta_{j,k_z,k'_z}(t) e^{-i(\omega_0+\omega_{k'})t} e^{ik'_z z_j} dk'_z\right)\right] \quad (2\text{-a})$$

$$i\dot{\beta}_j(t) = \int_{-\infty}^{+\infty} g_k\left[\left(\alpha_{k_z}(t)e^{i(\omega_0-\omega_k)t} e^{ik_z z_j}\right) + \left(\gamma_{k_z}(t)e^{-i(\omega_0+\omega_k)t} e^{ik_z z_{j'}}\right)\right] dk_z;\ (j' \neq j) \quad (2\text{-b})$$

$$i\dot{\gamma}_{k_z}(t) = g_k \sum_{\substack{j=1,2 \\ j \neq j'}} \beta_j(t) e^{-ik_z z_{j'}} e^{i(\omega_0+\omega_k)t} + 2\sum_{\substack{j=1,2 \\ j \neq j'}} \int_{-\infty}^{+\infty} g_{k'}\, \eta_{j,k_z,k'_z}(t) e^{i(\omega_0-\omega_{k'})t} e^{ik'_z z_j} dk'_z \quad (2\text{-c})$$

$$i\dot{\eta}_{j,k_z,k'_z}(t) = \dfrac{1}{2}\left[ g_{k'}\left(\alpha_{k_z}(t) e^{i(\omega_0+\omega_{k'})t} e^{-ik'_z z_j}\right) + g_{k'}\left(\gamma_{k_z}(t) e^{-i(\omega_0-\omega_{k'})t} e^{-ik'_z z_{j'}}\right) + (k_z \leftrightarrow k'_z)\right];\ (j' \neq j) \quad (2\text{-d})$$

These equations show that states with excitation number equal to one (e.g. $|a_1, a_2, 1_{k_z}\rangle$ and $|a_{j'\neq j}, b_j, 0\rangle$) are coupled through RWA coupling terms (operators $\hat{a}_{k_z}^\dagger \hat{\sigma}_j$ and $\hat{a}_{k_z} \hat{\sigma}_j^\dagger$) whereas states with excitation number of three (e.g. $|b_1, b_2, 1_{k_z}\rangle$ and $|b_j, 1_{k_z}, 1_{k'_z}\rangle$) are coupled respectively to $|a_{j'\neq j}, b_j, 0\rangle$ and $|a_1, a_2, 1_{k_z}\rangle$ because of non-RWA coupling terms (operators $\hat{a}_{k_z} \hat{\sigma}_j$ and $\hat{a}_{k'_z}^\dagger \hat{\sigma}_j^\dagger$). Finally, RWA coupling between highly excited states $|b_1, b_2, 1_{k_z}\rangle$ and $|b_j, 1_{k_z}, 1_{k'_z}\rangle$ also appears in (2-c) and (2-d).

### II-1 Atomic coupling:

The system of equations (2) can be considerably simplified because of the presence of a continuum of modes. In annex 1, we show that when $\omega_0, c/l \gg \Gamma, \Delta$, a Markovian approximation can be used leading to the fundamental equation for the amplitudes $\beta_j(t)$ ($j=1,2$):

$$\dot{\beta}_j(t) = S_{0,i}(t) - \Gamma \beta_j - M \beta_{j'\neq j}(t) \quad (3)$$

where $\Gamma = \dfrac{2\pi}{c}\dfrac{g_k^2 \omega_k}{\omega_0}$ is a relaxation constant term (independent of frequency $\omega_k$) and

$S_{0,j}(t) = -i\sqrt{\dfrac{\Gamma}{2\pi}} \int_{-\infty}^{+\infty} \sqrt{c\dfrac{\omega_0}{\omega_k}}\, \alpha_{k_z}(t \to -\infty) e^{i(\omega_0-\omega_k)t} e^{ik_z z_j}\, dk_z$ a source term due to the presence of an initial incident photon. Equation (3) exhibits also a third term that results from the coupling of the two atoms through the field and that involves a coupling parameter $M$ that is the sum of four contributions $M = \sum_{i=1}^{4} M_i$ corresponding to different quantums paths as it will be explained further:

$$M_1 = \frac{\Gamma \omega_0}{2\pi} \int_0^\infty \int_0^\infty \frac{e^{i(\omega_0 - \omega)\tau} e^{ik_z l}}{\omega} d\omega d\tau \quad (4\text{-a})$$

$$M_2 = \frac{\Gamma \omega_0}{2\pi} \int_0^\infty \int_0^\infty \frac{e^{-i(\omega_0 + \omega)\tau} e^{ik_z l}}{\omega} d\omega d\tau \quad (4\text{-b})$$

$$M_3 = M_1(l \leftrightarrow -l) = \frac{\Gamma \omega_0}{2\pi} \int_0^\infty \int_0^\infty \frac{e^{i(\omega_0 - \omega)\tau} e^{-ik_z l}}{\omega} d\omega d\tau \quad (4\text{-c})$$

$$M_4 = M_2(l \leftrightarrow -l) = \frac{\Gamma \omega_0}{2\pi} \int_0^\infty \int_0^\infty \frac{e^{-i(\omega_0 + \omega)\tau} e^{-ik_z l}}{\omega} d\omega d\tau \quad (4\text{-d})$$

Using the the mathematical relations:

$$\int_0^{+\infty} e^{i(\omega - \omega_0)T} dT = \pi \delta(\omega - \omega_0) + i\wp\left(\frac{1}{\omega - \omega_0}\right) \quad (5\text{-a})$$

$$\int_0^{+\infty} e^{i(\omega + \omega_0)T} dT = i\wp\left(\frac{1}{\omega + \omega_0}\right) \quad (5\text{-b})$$

where $\wp$ designs the Cauchy principal part of the integral, we obtain :

$$M_1 = \frac{\Gamma e^{ik_0 l}}{2} + \frac{i\Gamma \omega_0}{2\pi} \wp\left(\int_0^\infty \frac{e^{i\omega l/c}}{\omega_0 - \omega} \frac{d\omega}{\omega}\right) \quad (6\text{-a})$$

$$M_2 = -\frac{i\Gamma \omega_0}{2\pi} \int_0^\infty \frac{e^{i\omega l/c}}{\omega_0 + \omega} \frac{d\omega}{\omega} \quad (6\text{-b})$$

$$M_3 = \frac{\Gamma e^{-ik_0 l}}{2} + \frac{i\Gamma \omega_0}{2\pi} \wp\left(\int_0^\infty \frac{e^{-i\omega l/c}}{\omega_0 - \omega} \frac{d\omega}{\omega}\right) \quad (6\text{-c})$$

$$M_4 = -\frac{i\Gamma \omega_0}{2\pi} \int_0^\infty \frac{e^{-i\omega l/c}}{\omega_0 + \omega} \frac{d\omega}{\omega} \quad (6\text{-d})$$

The integrals appearing in (6) can be evaluated with the introduction of the sine ($Si$) and cosine ($Ci$) integral functions defined by $Ci(x) = -\int_x^\infty \frac{\cos t}{t} dt; Si(x) = \int_0^x \frac{\sin t}{t} dt$ [49]. For real arguments, these functions are even and odd respectively. The asymptotic values are $Ci(|x| \gg 1) = 0, Si(|x| \gg 1) = \pi/2$ and we have $Ci(0) = \infty$. We obtain the following relations for the coupling elements:

$$M_1 = \frac{\Gamma}{2} e^{i(\omega_0 l/c)} + \frac{\Gamma}{2\pi}\left[e^{i(\omega_0 l/c)}\left(Si(\omega_0 l/c) + \frac{\pi}{2} + i\, Ci(\omega_0 l/c)\right) - G_+\right] \quad \text{(7-a)}$$

$$M_2 = \frac{\Gamma}{2\pi}\left[e^{-i(\omega_0 l/c)}\left(Si(\omega_0 l/c) - \frac{\pi}{2} - iCi(\omega_0 l/c)\right) + G_+\right] \quad \text{(7-b)}$$

$$M_3 = \frac{\Gamma}{2} e^{-i(\omega_0 l/c)} + \frac{\Gamma}{2\pi}\left[-e^{-i(\omega_0 l/c)}\left(Si(\omega_0 l/c) + \frac{\pi}{2} - i\, Ci(\omega_0 l/c)\right) - G_-\right] \quad \text{(7-c)}$$

$$M_4 = \frac{\Gamma}{2\pi}\left[-e^{i(\omega_0 l/c)}\left(Si(\omega_0 l/c) - \frac{\pi}{2} + iCi(\omega_0 l/c)\right) + G_-\right] \quad \text{(7-d)}$$

$G_\pm$ is a constant given by $G_\pm = \pm\frac{\pi}{2} + iCi(\varepsilon l/c \to 0)$ and its imaginary part diverges ($\varepsilon$ is an artificial infrared frequency cutoff). This is not surprising since the atom-photon coupling is $g_k \propto \omega_k^{-1/2}$ and diverges also in the infrared. However, only the sum of these integrals is involved in the integral in (3) and the final coupling term $M = \sum_{i=1}^{4} M_i$ is convergent and given by:

$$M = \Gamma e^{ik_0 l} \quad \text{(8)}$$

The dependence of the coupling coefficient with the interatomic distance appears through the dephasing term $e^{ik_0 l}$. Thus, the coupling term doesn't decrease with the atomic separation in contrast with the free space situation. This is because in our situation (1-d waveguide with $l \gg d$), the propagating photons are confined along the inter atomic axis making the energy flux unchanged between atoms. This is in contrast with the free space where the emission of the photon with wavevectors out of the interatomic axis is allowed, reducing the photon exchange probability by $1/l$ and $1/l^3$ decreasing terms for isotropic and anisotropic emission respectively [31-33, 37]. Note that the infrared divergence is particular to the 1-d case where the state density is constant with the frequency $\omega_k$. In 3-d situation (free space), the state density ($\propto \omega_k^2$) compensates for the $g_k^2$ contribution ($\frac{1}{\omega_k}$) and one therefore deal with an ultraviolet divergence of the amplitudes [37-42, 50].

## II-2 Quantum paths

The field and the atomic dynamic can be also understood from equation (3) in terms of photon exchange between atoms. Moreover, we represent in figure 2 the paths corresponding to all terms of equation (3) stressing on the photon absorption and emission processes. We have considered the evolution of the excited state of the first atom for simplicity. The first term (source term) is represented schematically in (a) and corresponds to the situation where the incident photon is absorbed by the atom in the ground state leading to an increase of the excited state amplitude. The second term corresponds to (b) and represents the excited state relaxation because of the coupling to a continuum of photons. The photons are emitted in both reflected and transmitted directions. Note that these two schemes hold also when only a single atom is present. The other remaining terms correspond to the interaction between atoms. Case (c), representing the case where atom 2 relaxes and emits photons corresponds to the presence of RWA contributions. The photon emitted in the backward direction interacts with atom 1 leading to a modification of the excited state amplitude. This situation is associated with $M_1$ contribution to the coupling term. The case where the photon is emitted in the forward direction is associated with the $M_3$ contribution to the coupling term. Case (d) represents the situation corresponding to the presence of non RWA terms. The path corresponding to $M_4$ contribution represents the case where atom 1 emits a photon and transits to the excited states. The forward photon is then absorbed by atom 2 that relaxes to the ground state. The $M_2$ contribution is the same process than $M_4$ but with the absorption of the backward photon by atom 2. It's worthy here to notice that although the photon emitted in the forward direction fly away from atom 1 ($M_3$ contribution) or 2 ($M_2$ contribution) it can interact with these atoms because the coupling diverges for long wavelengths explaining for non vanishing contributions of these terms.

## II-3 Real and virtual photons interplay:

The coupling term and paths associated with $M_j$ contributions can also be related to the virtual and real characters of the photons involved in the process. According to the common signification of these expressions, the real photons are those created in resonant physical processes that conserve the bare energy (e.g. energy without atom-radiation coupling) whereas virtual photons are the ones that are created in non-resonant processes that do not conserve this energy [51]. In our situation, real photon corresponds to the resonant photon ($\omega = \omega_0$) involved in RWA contributions ($M_1$ and $M_3$). These observe bare-energy conservation during the whole exchange process. Virtual photons are the others non-resonant photons ($\omega \neq \omega_0$) involved in RWA processes ($M_1$ and $M_3$) and all photons involved in non-RWA processes ($M_2$ and $M_4$) that obviously violate bare-energy conservation in intermediate states of the system. Both real and virtual photons can modify the dynamics of the quantum system but in different manner as we see next. We refer back to formulas (6) to understand the photon contribution. It's worthy to notice that real photon contribution originates from the Dirac function part in $M_1$ (term $\frac{\Gamma e^{ik_0 l}}{2}$) and $M_3$ (term $\frac{\Gamma e^{-ik_0 l}}{2}$). Their sum originates in the presence of the real part $-\Gamma \cos(k_0 l)$ of the coupling term in (6-a) and (6-b). Virtual photons involve the remaining terms and contribute to the imaginary part of the coupling $-i\Gamma \sin(k_0 l)$. Thus, the real part expresses population modifications for the atoms (e.g. transitions) whereas the imaginary part is associated with a frequency shift of the atomic resonances.

**II-3 Importance of non-RWA contribution:**

The expression of $M_i$ (7) show clearly that non-RWA contributions can be as important as RWA ones although the inequality $\Gamma << \omega_0$ is assumed to use the adiabatic approximation. To strengthen the importance of non-RWA terms and to find situations where it would be possible to restrict the calculations to RWA, we consider many instructive situations.

If RWA is performed, the coupling term between atoms would be $M^{(RWA)} = M_1 + M_3$ and is given by :

$$M^{(RWA)} = \Gamma \cos(\omega_0 l/c) + i\frac{\Gamma}{\pi}\left(\sin(\omega_0 l/c)\left(Si(\omega_0 l/c) + \frac{\pi}{2}\right) + \cos(\omega_0 l/c)Ci(\omega_0 l/c) - Ci(\varepsilon l/c \to 0)\right) \quad (9)$$

Only the real part is correct with respect to the true value of $M$. The imaginary part diverges. The non-RWA contributions are thus necessary for convergence of the coupling parameter *(without the need of any frequency cutoff)*. However, it's worthy to notice that this is not the only role these terms play. Indeed, the addition of non-RWA parts $M_2, M_4$ *also introduces an* *additional partial shift* $\frac{\Gamma}{\pi}\left[\sin(\omega_0 l/c)\left(-Si(\omega_0 l/c)+\frac{\pi}{2}\right)-\cos(\omega_0 l/c)Ci(\omega_0 l/c)\right]$ *giving the true shift* $\Gamma\sin(\omega_0 l/c)$ *in M.*

Another interesting situation is the case where $\omega_0 l/c \gg 1$. In this case, $Ci(\omega_0 l/c) \simeq 0$; $Si(\omega_0 l/c) \simeq \pi/2$. The RWA give the right result only if an infrared frequency cutoff $\varepsilon$ is introduced such as $\varepsilon \gg c/l$. In this case $Ci(\varepsilon l/c) \simeq 0$ and $M^{(RWA)} \simeq M = \Gamma e^{i(\omega_0 l/c)}$.

The correct model doesn't need for any frequency cut-off as said before and a physical interpretation involving virtual photons can be given when $\omega_0 l/c \gg 1$. Indeed, in this case we can separate the non-resonant contribution (e.g. the integral) in (6-a and c) into two parts, one corresponding to photons nearly resonant with frequencies $\omega$ located in a domain $\delta \geq c/l$ around $\omega_0$ (with $\omega_0 \gg \delta$) and another part with the remaining photons. In this situation, considering $M_1$ contribution ($M_3$ respectively), we have $\frac{i\Gamma\omega_0}{2\pi}\wp\left(\int_{\omega_0-\delta/2}^{\omega_0+\delta/2}\frac{e^{i\omega l/c}}{\omega_0-\omega}\frac{d\omega}{\omega}\right) \simeq \frac{\Gamma e^{ik_0 l}}{2}$ ($\frac{i}{2\pi}\wp\left(\int_{\omega_0-\delta/2}^{\omega_0+\delta/2}\frac{e^{-i\omega l/c}}{\omega_0-\omega}\frac{d\omega}{\omega}\right) \simeq -\frac{\Gamma e^{-ik_0 l}}{2}$ respectively). The sum of these contributions gives rise to the imaginary part of the coupling $i\Gamma\sin(k_0 l)$. In other words, only nearly resonant photons contribute to the atomic coupling. The role of the remaining part (highly non resonant photons in $M_1$ and $M_3$) is to annihilate the non-RWA photons contributions ($M_2$ and $M_4$). This result is in line with the (undesired) non-causal character of the interaction associated with these photons. Indeed, due to time-energy incertitude energy, non-RWA photons are present for a time $|\omega+\omega_0|^{-1} \leq \omega_0^{-1}$. So, they should not exceed a travel distance of about $\lambda_0 = c/\omega_0$ in accordance with causality (finite *c*). However, the divergence of the interaction parameter in the infrared domain (responsible for the $1/\omega$ term in the integrals (6-7)) leads to an efficient interaction between atoms even if $l > \lambda_0$, thus violating causality. These contributions have necessarily to be compensated in the expression of any measurable physical quantity to fulfill causality principle.

A frequently situation considered also is the case where RWA is used and the frequency variation of the coupling $g_\omega$ is neglected (e.g. $g_\omega(\omega_k) = g_\omega(\omega_0)$). In this case, using our notations, the same calculations lead to substitution of $1/\omega$ by $1/\omega_0$ in the integrals of equation (4-a, c and 6-a, c) giving a coupling $M^{(RWA)}_{g_\omega cte}$:

$$M_{g_\omega cte}^{(RWA)} = \Gamma \cos(\omega_0 l/c) + i\frac{\Gamma}{\pi}\left[\sin(\omega_0 l/c)\left(Si(\omega_0 l/c) + \frac{\pi}{2}\right) + \cos(\omega_0 l/c)Ci(\omega_0 l/c)\right] \quad (10)$$

This result corresponds to $M^{(RWA)}$ with the remove of the diverging term. However, except the situation where $\omega_0 l/c \gg 1$ this model is not suitable to recover the true coupling parameter $M = \Gamma e^{i(\omega_0 l/c)}$.

Finally, another procedure used in some works [52], is to perform RWA with the extension of integration in the coupling parameter to negative frequencies (e.g. in (6-a, c) the integration is performed from $-\infty$ to $+\infty$). This gives a coupling parameter $M_{-\infty,\infty}^{(RWA)} = \Gamma e^{i(\omega_0 l/c)}$ that exactly matches with the true coupling $M$. This procedure was introduced in previous works to preserve causality in photodetection processes [53] and found here its justification in our case.

### III- <u>Field behavior</u> :

#### III-a Photoelectric signal:

The field behaviour is modified by the interaction with the atoms. An important feature already mentioned in the one atom case [2, 18, 23-24] is the reflection of the resonant frequency of the field. This property leads to a transmission of a electric field that distorts temporally such as its algebraic area vanishes (pulse-area theorem [44]). This important feature is the key point to understand the field dynamics. Here, we show that the presence of non-RWA terms is necessary to preserve this feature and is the consequence of the compensation between parts of virtual photons contributions.

We consider the mean field intensity at a photodetector located at a distance $z$ from atom 1. We assume that the photodetector is fast enough to resolve the temporal variation of the entering field. In Glauber theory of photodetection (RWA done in the detector), the photodetector signal -in Coulomb jauge- is then given by $I(t,z) = s\langle\psi(t)|\hat{A}^{(-)}(z)\,\hat{A}^{(+)}(z)|\psi(t)\rangle$ ($s$ a constant set equal to 1 for simplicity). In this notation, $\hat{A}^{(\pm)} = \int_{-\infty}^{+\infty} B(\omega_k) dk_z \frac{\varepsilon_k}{\omega_k} \hat{a}_{k_z} e^{\pm ik_z z}$ is the positive (respectively negative) frequency part of the potential vector field operator, $\varepsilon_k$ the vacuum electric field and $|\psi\rangle$ is the wave function. We introduce in these expressions $B(\omega_k)$ the spectral acceptance of the detector defined as $B(\omega_k) = 1$ for $\omega_1 < \omega_k < \omega_2$ and $B(\omega_k) = 0$ elsewhere, $\omega_{2,1} = \omega_0 \pm \frac{\Delta_0}{2}$, $\Delta_0 = \omega_2 - \omega_1$ is the detector spectral bandwidth satisfying $\Delta_0 \gg \Delta, \Gamma$ to ensure the spectral collection of all emitted photons. We also assume that $|z - z_j| \gg c\Gamma^{-1}, c\Delta^{-1}$ to ensure that the field emission is complete before its entry in the photodetector. Using the above definitions and expression (1) of the wavefunction, we find the following expression for the mean field intensity $I(t,z) = I_1 + I_2 + I_3$ with $I_1(t,z) = |A_{eff}(t,z)|^2$, $A_{eff}$ given by :

$$A_{eff}(t,z) = \int_{-\infty}^{+\infty} \left(\frac{\varepsilon_k}{\omega_k}\right) B(\omega_k) \alpha_{k_z}(t) \, e^{-i\omega_k(t - sign(k_z)z/c)} dk_z \tag{11}$$

and

$$I_2(t,z) = \left|\int_{-\infty}^{+\infty} \left(\frac{\varepsilon_k}{\omega_k}\right) B(\omega_k) \gamma_{k_z}(t) \, e^{-i\omega_k(t - sign(k_z)z/c)} dk_z\right|^2 \tag{12-a}$$

$$I_3(t,z) = 2\sum_{j=1}^{2} \int_{-\infty}^{+\infty} dk_z \left|\int_{-\infty}^{+\infty} dk'_z \left(\frac{\varepsilon_{k'}}{\omega_{k'}}\right) B(\omega_{k'}) \eta_{j,k_z,k'_z}(t) \, e^{-i\omega_{k'}(t - sign(k'_z)z/c)}\right|^2 \tag{12-b}$$

$I_1$ represents the intensity due to the incident field and the field radiated through RWA processes. The intensity expression is similar to that obtained in the classical regime with an effective potential vector field $A_{eff}$. $I_2$ and $I_3$ are associated with fields radiated trough non-RWA processes and are thus *exclusively due* to virtual photons. In annex 2, we show that $I_2, I_3 \simeq 0$ as long as $|z| \gg c/\omega_1$ and $z < 0$, or $|z - l| \gg c/\omega_1$ and $z > 0$ (conditions that are automatically fulfilled in our situation with $|z - z_j| \gg c\Gamma^{-1}, c\Delta^{-1}$). The vanishing of $I_2$ and $I_3$ can be understood from the fact that these intensities are associated with (non-RWA) virtual photons that are located within a wavelength from the atoms. As $c/\omega_1$ corresponds to the maximum wavelength accepted by the detector, none of these virtual photons influence the photodetection process when the detector is located at a larger distance from the atoms. If the detector is in the near field regime ($|z - z_j| \leq \lambda_1 = c/\omega_1$), a non-RWA treatment of the whole interaction (atoms + photodetector) is needed [53, 54]. The effective field $A_{eff}$ in relation (11) can be related to the population amplitude $\beta_j$ of excited states. In annex 3, we show that in the limit of Markovian approximation, the effective field can be decomposed in three propagating parts as follows:

$$A_{eff}(t,z) = h(-z)A_{inc}(t-z/c)e^{-i\omega_0(t-z/c)} + h(-z)A_{refl}(t+z/c)e^{-i\omega_0(t+z/c)} + h(z-l)A_{trans}(t-z/c)e^{-i\omega_0(t-z/c)} \quad (13)$$

With :

$$A_{inc}(t-z/c) = \int_0^\infty B\frac{\varepsilon_k}{\omega_k}\left(\alpha_{k_z}(-\infty)e^{-i(\omega_k-\omega_0)(t-z/c)}\right)dk_z \quad \text{(14-a)}$$

$$A_{trans}(t-z/c) = A_{incid}(t-z/c) - i\frac{g_k\varepsilon_k}{c}\frac{2\pi}{\omega_0}\sum_{j=1}^2 e^{-i\frac{\omega_0}{c}z_j}\beta_j(t-(z-z_j)/c) \quad \text{(14-b)}$$

$$A_{refl}(t+z/c) = -i\frac{g_k\varepsilon_k}{c}\frac{2\pi}{\omega_0}\sum_{j=1}^2 e^{i\frac{\omega_0}{c}z_j}\beta_j(t+(z-z_j)/c) \quad \text{(14-c)}$$

$A_{inc}(t,z)$, $A_{refl}(t,z)$ and $A_{trans}(t,z)$ are respectively the incident, reflected and transmitted electric wavepackets ($h(z)$ is the Heaviside function). Finally, another interesting quantity used in our investigations is the spectral distribution of the field $\tilde{I}_\alpha(\omega) = \left|\tilde{A}_\alpha(\omega)\right|^2$ that gives the energy distribution of the corresponding photons ($\alpha$ stands for incident, transmitted and reflected and $\tilde{A}_\alpha(\omega) = \int_{-\infty}^{+\infty} A_\alpha(\tau)e^{i(\omega-\omega_0)\tau}d\tau^2$).

### III-2 Transmitted and reflected wavepackets : Pulse-area theorem.

We establish in this section that the algebraic pulse-area of the transmitted and reflected pulse obey to strict conditions. Indeed, the transmitted potential field is given by (14-b) and for $z = z_j$ ($j=1,2$), we have

$$A_{trans}(t-z_j/c) = A_{incid}(t-z_j/c) - i\frac{g_k\varepsilon_k}{c}\frac{2\pi}{\omega_0}\sum_{j'=1}^2 e^{-i\frac{\omega_0}{c}z_{j'}}\beta_{j'}\left(t-(z_j-z_{j'})/c\right) \quad \text{and}$$

$$A_{refl}(t+z_j/c) = -i\frac{g_k\varepsilon_k}{c}\frac{2\pi}{\omega_0}\sum_{j'=1}^2 e^{i\frac{\omega_0}{c}z_{j'}}\beta_{j'}(t+(z_j-z_{j'})/c).$$ Introducing the constant

$G_{0j} = \frac{g_k\omega_k}{\varepsilon_k}e^{ik_0z_j}$ (independent of $\omega_k$) and remembering that we are working within the Markovian approximation $\beta_j(t\pm l/c) \simeq \beta_j(t)$ (and $A_{incid}(t-l/c) \simeq A_{incid}(t)$), we found that the variation $\dot{\beta}_j$ of the excited state population given by (3) is directly related to the contribution of propagating fields through the following relations :

$$i\dot{\beta}_1(t) = G_{01}\left(A_{inc}(t) + A_{refl}(t)\right) \quad (15\text{-a})$$

$$i\dot{\beta}_2(t) = G_{02} A_{trans}(t) \quad (15\text{-b})$$

This an important property: Although virtual photons are taken into account, only the propagating –causal- fields (incident, reflected and transmitted) evaluated at the atomic position modify the dynamics of the corresponding population. This is the consequence of the interference of the contribution of virtual photons due to two-photon terms (non-RWA terms) with the contribution that originates from virtual photons with energies located outside a narrow bandwidth around the resonant frequency (RWA contribution) as discussed in §II-3. We define the pulse area as $S_i = \int_{-\infty}^{+\infty} A_i(\tau = t - z/c) d\tau$ ($i = inc, trans, refl$). Integration of equation (12-b) turns into $S_{trans} = i\, G_{02}\left(\beta_2(t \to +\infty) - \beta_2(t \to -\infty)\right)$. The asymptotic behaviour of $\beta_j$ (j=1,2) can be obtained by deduced from equation (3). Moreover, because of the presence of the relaxation term, we have $\beta_j(t \to +\infty) = 0$ meaning that the atoms come back to the initial ground states after the end of photon scattering process. Because initially $\beta_j(t \to -\infty) = 0$ the transmitted pulse area vanishes e.g. $S_{trans} = \int_{-\infty}^{+\infty} A_{trans}(\tau = t - z/c) d\tau = 0$. Similarly, we have $\int_{-\infty}^{+\infty}\left(A_{inc} + A_{refl}\right)(\tau = t - z/c) d\tau = 0$. We finally obtain the following important results:

$$S_{trans} = \int_{-\infty}^{+\infty} A_{trans}(\tau = t - z/c) d\tau = 0 \quad (16\text{-a})$$

$$S_{refl} = \int_{-\infty}^{+\infty} A_{refl}(\tau = t - z/c) d\tau = -\int_{-\infty}^{+\infty} A_{inc}(\tau = t - z/c) d\tau = -S_{inc} \quad (16\text{-b})$$

Note that this result is valid whatever is the coupling $\Gamma$, the pulse width $\Delta$ and the distance $l$ between the atoms. The pulse area can be identified in the spectral domain with the spectrum at resonance. Thus, equations (16) means that the central frequency is always *totally* reflected. Moreover, the atoms radiates in both backward and forward directions but for the resonance frequency the interference between radiated fields is always destructive (constructive) in the forward (backward) direction.

An important remark has to be done at this level. The pulse-area theorem holds only when equations (15) are valid. As noticed above, this is the consequence of the interference between the contributions of virtual photons due to non-RWA terms and those of non-resonant RWA terms in the expression of fields. Thus only RWA 1-d models that recover the correct shift can recover this feature (see §II-3).

These results are illustrated in figure 3 where the temporal and spectral profiles of the transmitted pulse is plotted for a fixed value of the distance $l$ such as both virtual and real photons are involved ($k_0 l = \pi/4$) and for three increasing values of the ratio $\Gamma/\Delta$. In all cases, both the (a) real and (b) imaginary parts of the field exhibit both positive and negative parts to ensure the vanishing of the pulse area. The distortion of the pulse increases also with the coupling parameter and oscillations appear with a characteristic time that becomes smaller than the initial pulse duration for large values of the coupling. This distortion is also accompanied with a significant decrease of the total amplitude. This is in line with equation (15). When $\Gamma$ increases, $G_{02}$ decreases and $A_{trans}(t-l/c) = -i G_{02}^{-1} \dot{\beta}_2(t)$ vanishes as a consequence. In the same manner, we obtain from (15) that $A_{refl}(t) \simeq -A_{inc}(t)$ for large coupling parameter $\Gamma/\Delta$. In (c), we represent the spectrum of the transmitted field for corresponding values of the coupling parameter. We see that the central frequency is never transmitted and is hence totally reflected. The spectrum exhibits also a profound dip whose width increases with the coupling parameter.

## IV- **Conclusion:**

A detailed study of the scattering of an incident photon wave-packet by two atoms in one dimensional waveguide has been presented. We clarify the role and importance of non-RWA terms to account correctly for virtual photons contributions. Moreover, we have shown that a subtle interplay between parts of virtual photons contribution leads to strong constraints on the pulse-are of temporal envelopes. This study show that virtual photons can lead to substantial –quantitative- modification of both atomic and radiated fields in the one dimensional waveguide, in line with the 3-d case. Extension of this work to an array of N atoms and for non-Markovian case is a natural perspective. Moreover, the interpretation in terms of temporal behavior for the fields developed here turns out to be a useful concept to understand shaping effects. This constitutes as a first step in the manipulation of photon wave packet characteristics thus adding a new control parameter –the shape- for the transport of flying qubits.

**Annex 1**: Fundamental equation for the populations

The set of equation (2) can be simplified so as a simple equation can be obtained for the excited state populations $\beta_j(t)$. First, the relevant parts of the amplitudes of highly excited states involved in (2-a) and (2-b) can be obtained by integrating formally equations (2-c) and (2-d) using the approximation $\int_{-\infty}^{t} dt' f(t') e^{i(\omega_0+\omega_k)(t'-t)} \simeq f(t) \int_{-\infty}^{t} dt' e^{i(\omega_0+\omega_k)(t'-t)} \simeq \frac{f(t)}{i(\omega_0+\omega_k)}$ ($f(t) = \alpha_{k_z}(t)$ or $\beta_j(t)$). We obtain :

$$\gamma_{k_z}(t) \simeq -\frac{g_k}{\omega_0+\omega_k} \sum_{\substack{j=1,2 \\ j\neq j'}} \beta_j(t) e^{-ik_z z_{j'}} e^{i(\omega_0+\omega_k)t} \quad \text{(A1-1a)}$$

$$\eta_{j,k_z,k'_z}(t) \simeq -\frac{g_k}{\omega_0+\omega_{k'}} \left( \alpha_{k_z}(t) e^{i(\omega_0+\omega_{k'})t} e^{-ik'_z z_j} \right) \quad \text{(A1-1b)}$$

Injecting equation (A1-1b) into (2-a). We obtain the following equation for $\alpha_{k_z}(t)$:

$$i\dot{\alpha}_{k_z}(t) \simeq \left( g_k \sum_{j=1,2} \left( \beta_j(t) e^{-i(\omega_0-\omega_k)t} e^{-ik_z z_j} \right) \right) - \left( \alpha_{k_z}(t) \int \frac{2g_{k'}^2}{\omega_0+\omega_{k'}} dk'_z \right) \quad \text{(A1-2)}$$

The inclusion of two-photon states $\left| \beta_j, 1_{k_z}, 1_{k'_z} \right\rangle$ in the dynamics of the system leads to a shift of the ground levels (factor 2 in (A1-2) appears because of the summation over $j$. This is the usual Lamb-shift due to the vacuum because of the emission-absorption cycles of virtual photons by atoms in the ground states. We next rewrite $\gamma_{k_z}$ as $\gamma_{k_z}(t) \simeq -\frac{g_k}{\omega_0+\omega_k} \beta_j(t) e^{-ik_z z_{j'}} e^{i(\omega_0+\omega_k)t} - ig_k \int_{-\infty}^{t} \beta_{j'}(t') e^{-ik_z z_j} e^{i(\omega_0+\omega_k)t'}$ (e.g. we take the part of $\gamma_{k_z}$ in $\beta_j$ (from A1-1a) and maintain the integral form $\beta_{j'}$ (from 2-c). Injecting this expression in (2-b), we obtain:

$$i\dot{\beta}_j(t) = \int dk_z \left[ g_k \left( \alpha_{k_z}(t) e^{i(\omega_0-\omega_k)t} e^{ik_z z_j} \right) + \left( -ig_k^2 \int_{-\infty}^{t} \beta_{j'}(t') e^{ik_z(z_{j'}-z_j)} e^{i(\omega_0+\omega_k)(t'-t)} dt' \right) \right] \quad \text{(A1-3)}$$
$$- \left( \beta_j(t) \int \frac{g_k^2}{\omega_0+\omega_k} dk_z \right); j' \neq j$$

The excited state exhibits also a shift that is half that of the ground state. Shifting the total energy of the system by the amount of ground state shift, integrating formally (A1-2), we obtain:

$$\alpha_{k_z}(t) = \alpha_{k_z}(t \to -\infty) - i\left(g_k \, e^{-ik_z z_j} \sum_{j=1,2} \int_{-\infty}^{t} dt' \left(\beta_j(t') e^{-i(\omega_0 - \omega_k)t'}\right)\right) \quad \text{(A1-4)}$$

Using this expression in the equation of evolution (A1-3) and the adiabatic elimination of the continuum [55] for the first term of the integral $\int g_k^2 dk_z \left(\int_{-\infty}^{t} \beta_j(t') e^{-i(\omega_0 - \omega_k)(t'-t)} dt'\right) \simeq \beta_j(t) \left(\int_{-\infty}^{t} \left(\int g_k^2 dk_z e^{-i(\omega_0 - \omega_k)(t'-t)}\right) dt'\right)$, we obtain the following fundamental equation for $\beta_j(t)$:

$$\dot{\beta}_j(t) = S_{0,j}(t) - (\Gamma - i\delta_0)\beta_j(t) - \Gamma\omega_0 \left(\int_{-\infty}^{t} \beta_{j'}(t') \bar{M}(t'-t) dt'\right) \quad \text{(A1-5)}$$

with $\Gamma = 2\pi g_0^2$ ($g_0 = \frac{g_k}{\sqrt{c}} \sqrt{\omega_k / \omega_0}$) and

$S_{0,j}(t) = -i\sqrt{\frac{\Gamma}{2\pi}} \int_{-\infty}^{+\infty} \sqrt{c\frac{\omega_0}{\omega_k}} \, \alpha_{k_z}(t \to -\infty) e^{i(\omega_0 - \omega_k)t} e^{ik_z z_j} dk_z$. $\bar{M} = \sum_{i=1}^{4} \bar{M}_i$ is the memory function and the $\bar{M}_i$ are defined by the relations:

$$\bar{M}_1(t-t') = \frac{1}{2\pi} \int_0^{\infty} \frac{e^{i(\omega_0 - \omega_k)(t-t')} \, e^{ik_z l}}{\omega_k} d\omega_k \quad \text{(A1-6-a)}$$

$$\bar{M}_2(t-t') = \frac{1}{2\pi} \int_0^{\infty} \frac{e^{-i(\omega_0 + \omega_k)(t-t')} \, e^{ik_z l}}{\omega_k} d\omega_k \quad \text{(A1-6-b)}$$

$$\bar{M}_3(t-t') = M_1(t-t', l \leftrightarrow -l) = \frac{1}{2\pi} \int_0^{\infty} \frac{e^{i(\omega_0 - \omega_k)(t-t')} \, e^{-ik_z l}}{\omega_k} d\omega_k \quad \text{(A1-6-c)}$$

$$\bar{M}_4(t-t') = M_2(t-t', l \leftrightarrow -l) = \frac{1}{2\pi} \int_0^{\infty} \frac{e^{-i(\omega_0 + \omega_k)(t-t')} \, e^{-ik_z l}}{\omega_k} d\omega_k \quad \text{(A1-6-d)}$$

$\delta_0 = \int \frac{g_k^2}{\omega_0 + \omega_k} dk_z + \wp\left(\int \frac{g_k^2}{\omega_k - \omega_0} dk_z\right)$ is the resultant shift of the excited state and can be incorporated in the definition of the transition frequency $\omega_0$. An important case is the Markovian situation where the atoms are close enough so that the interaction (exchange of photons) can be considered as instantaneous compared to the atomic dynamics [56]. This is the case when the photon time of flight $l/c$ and the resonant period ($\frac{2\pi}{\omega_0}$) are smaller than the time characteristics of population amplitudes $\beta_j$ that are $\Gamma^{-1}$ and $\Delta^{-1}$. This is obtained for $l, \lambda_0 \ll c\Gamma^{-1}, c\Delta^{-1}$ (but $l < \lambda_0$ or $l > \lambda_0$ allowed). In this case, we can set $\beta_j(t') \simeq \beta_j(t)$ in the integral appearing in (A1-5). We then obtain the equation:

$$\dot{\beta}_j(t) = S_{0,i}(t) - \Gamma \beta_j - M \beta_{j' \neq j}(t) \tag{A1-7}$$

with $M = \int_{-\infty}^{t} \bar{M}(t-t')dt'; M_i = \int_{-∞}^{t} \bar{M}_i(t-t')dt'$.

**Annex 2**: Contribution of Non-RWA photons to the photoelectric signal

We consider $I_2(t,z) = \left| \int_{-\infty}^{+\infty} \left( \frac{\varepsilon_k}{\omega_k} \right) B(\omega_k) \gamma_{k_z}(t) e^{-i\omega_k(t-\text{sign}(k_z)z/c)} dk_z \right|^2$. We integrate formally equation (2-c) and insert it in the expression of $I_2(t,z)$. Using the adiabatic elimination of the continuum technic we obtain:

$$I_2(t,z) = \left| 2 \sum_{j=1}^{2} \beta_j(t) \int_0^{\infty} d(\omega_k/c) B(\omega_k) \left( \frac{\varepsilon_k g_k}{\omega_k (\omega_0 + \omega_k)} \right) \cos\left(\omega_k |z-z_j|/c\right) \right|^2 \quad (A2\text{-}1)$$

The integration over $\omega_k$ can be done analytically since $\varepsilon_k g_k$ is constant. We have :

$$\int_0^{\infty} d(\omega_k/c) B(\omega_k) \left( \frac{\cos\left(\omega_k |z-z_j|/c\right)}{\omega_k (\omega_0 + \omega_k)} \right) = f(\omega_2, \omega_0, a) - f(\omega_1, \omega_0, a) \quad (A2\text{-}2)$$

where :

$$f(\omega, \omega_0, a) = \frac{1}{c\omega_0} \left( -\cos(\omega_0 a) Ci((\omega+\omega_0)a) + Ci(\omega a) - \sin(\omega_0 a) Si((\omega+\omega_0)a) \right) \quad (A2\text{-}3)$$

With $a = |z-z_j|/c$. $Ci$ and $Si$ are the cosine and sine integral functions respectively [49]. For real arguments, these functions are even and odd respectively and the asymptotic values are $Ci(|x| \gg 1) = 0, Si(|x| \gg 1) = \pi/2$. From these properties it follows that $f(\omega_2, \omega_0, a) - f(\omega_1, \omega_0, a)$ vanishes as long as $\omega_1 |z-z_{j'}|/c \gg 1$ (and thus $\omega_2 |z-z_{j'}|/c \gg 1$). In this case, the intensity $I_2(t,z)$ vanishes as a result.

We consider now $I_3(t,z) = \sum_{j=1}^{2} \int_{-\infty}^{+\infty} dk_z \left| \int_{-\infty}^{+\infty} dk'_z \left( \frac{\varepsilon_{k'}}{\omega_{k'}} \right) B(\omega_{k'}) \eta_{j,k_z,k'_z}(t) e^{-i\omega_{k'}(t-\text{sign}(k'_z)z/c)} \right|^2$.

We calculate this expression by using equation (A1-1b), inject it in the above expression of $I_3$. We obtain $I_3(t,z) = \sum_{j=1}^{2} \int_{-\infty}^{+\infty} dk_z |D_{k_z,j}|^2$ with :

$$D_{k_z} = -\sqrt{2}\alpha_{k_z}(t)\int_{\omega_1}^{\omega_2} d\omega_{k'} \left(\frac{g_{k'}\varepsilon_{k'}}{c\omega_{k'}(\omega_0+\omega_{k'})}\right) B(\omega_{k'})\cos\left(\omega_{k'}(z-z_j)\right) \qquad \text{(A2-4)}$$

$$-i\frac{\sqrt{2}g_k e^{i(\omega_0+\omega_k)t}}{2(\omega_0+\omega_k)} e^{-ik_z z_j} \int_{-\infty}^{+\infty} dk_z' \left(\frac{\varepsilon_{k'}}{\omega_{k'}}\right) B(\omega_{k'})\alpha_{k_z'}(t) e^{i(k_z' z-\omega_{k'} t)}$$

If $\omega_1 |z-z_j|/c \gg 1$ the cosine term in the first integral in (A2-4) strongly oscillates and the corresponding integral vanishes. This can also be explicitly demonstrated using relations (A2-2 and A2-3). In the second term in (A2-5), we recognize the effective field $A_{eff}(t,z)$ given in (11). We then obtain the following expression for $I_3$:

$$I_3 \simeq I_1 \int_{-\infty}^{+\infty} \frac{g_k^2}{2(\omega_0+\omega_k)^2} dk_z \qquad \text{(A2-5)}$$

Using the expression of $g_k = \sqrt{\frac{\Gamma c\omega_0}{2\pi\omega_k}}$ and the relation $\int_\varepsilon^\infty dx/\left(x(1+x)^2\right) \simeq \ln(\varepsilon)$, we have $I_3 \simeq \sqrt{\frac{1}{2\pi}\frac{\Gamma}{\omega_0}}\ln(\omega_0/\omega_c) I_1$ where $\omega_c$ is a low frequency cutoff. Because $\Gamma/\omega_0 \ll 1$ and the slowly variation of the logarithmic term, $I_3 \ll I_1$. This achieve the demonstration that $I_2, I_3 \simeq 0$ for $\omega_1 |z-z_j|/c \gg 1$.

**Annex 3**: Relation between propagating fields and population amplitudes.

We consider the effective field $A_{eff}(t,z) = \int_{-\infty}^{+\infty} (\varepsilon_k/\omega_k) B(\omega_k) \alpha_{k_z}(t) e^{-i\omega_k(t-sign(k_z)z/c)} dk_z$.

Let's consider first the situation $z < 0$. The integration over $k_z$ can be separated into two integrals with $[0,\infty]$ and $[-\infty, 0]$ intervals respectively, we can rewrite $A_{eff}(t,z)$ as:

$$A_{eff}(t,z) = A_-(t,z)e^{-i\omega_0(t-z/c)} + A_+(t,z)e^{-i\omega_0(t+z/c)} \tag{A3-1}$$

with

$$A_-(t,z) = \int_0^\infty B(\omega_k) \frac{\varepsilon_k}{\omega_k} \left( \alpha_{k_z}(t) e^{-i(\omega_k-\omega_0)(t-z/c)} \right) dk_z \tag{A3-2a}$$

$$A_+(t,z) = \int_0^\infty B(\omega_k) \frac{\varepsilon_k}{\omega_k} \alpha_{-k_z}(t) e^{-i(\omega_k-\omega_0)(t+z/c)} dk_z \tag{A3-2b}$$

We show now that the first reduces to the incident one and the second term corresponds to the reflected wavepacket in situations where $\omega_1|z|/c \gg 1$ and $\Delta_0|z|/c \gg 1$ respectively. Indeed, using equation (A1-4), we have:

$$A_-(t,z) = \int_0^\infty B(\omega_k) \frac{\varepsilon_k}{\omega_k} \left( \alpha_{k_z}(t \to -\infty) e^{-i(\omega_k-\omega_0)(t-z/c)} \right) dk_z + \\ \int_0^\infty -iB(\omega_k) \frac{\varepsilon_k g_k}{\omega_k} e^{-ik_z z_j} \left( \left( \sum_{j=1,2} \int_{-\infty}^t dt' \left( \beta_j(t')e^{-i(\omega_0-\omega_k)t'} \right) \right) e^{i(\omega_0-\omega_k)(t-z/c)} \right) dk_z \tag{A3-3}$$

The next step is to show that the second integral in (A3-3) vanishes for $\omega_1|z|/c \gg 1$. Integration over $\omega_k$ is done first and we deals with the following integral: $\int_{\omega_1}^{\omega_2} \frac{e^{-i\omega_k T}}{\omega_k} dk_z = g(\omega_2) - g(\omega_1)$ with $T = t - t' - (z-z_j)/c$ and $g(\omega) = Ci(\omega T) - i\,Si(\omega T)$. $Ci$ and $Si$ are the cosine and sine integrals functions respectively [49]. The minimum value for $T$ is obtained for $t'=t$ and is $-(z-z_j)/c$ $(>0)$. Using the asymptotic values of the $Ci$ and $Si$ functions, we find that the integral vanishes as long as $\omega_1|z|/c \gg 1$. Thus, the amplitude $A_-(t,z)$ reduces to the incident wavepacket:

$$A_-(t,z) = A_{inc}(t - z/c) \tag{A3-4}$$

The last step is to show that the negative wavelength contribute (in A3-2b) to a reflected wave packet e.g. a wavepacket propagating with a $t + z/c$ dependence. We use expression (A1-4) for $\alpha_{-k_z}(t)$ and perform the adiabatic elimination of the continuum technic. Reminding the initial condition $\alpha_{-k_z}(t \to -\infty) = 0$ (no incident wavepacket coming from $z > 0$) and using relation (5-a), we obtain:

$$A_+(t,z) = -i\frac{g_k \varepsilon_k}{c} \sum_{j=1}^{2} \beta_j(t+(z-z_j)/c) \, e^{i\frac{\omega_0}{c}z_j} \left( \frac{\pi}{\omega_0} - i\wp \int B(\omega_k) d\omega_k \frac{e^{-i\frac{(\omega-\omega_0)}{c}(z-z_j)}}{\omega_k(\omega_k-\omega_0)} \right) \quad \text{(A3-5)}$$

The radiated field in this expression can be further simplified using the relation

$$\wp \int B(\omega_k) d\omega_k \frac{e^{-i\frac{\omega}{c}(z-z_j)}}{\omega_k(\omega_k-\omega_0)} = f_+(\omega_2, -\omega_0, -(z-z_j)/c) - f_+(\omega_1, -\omega_0, -(z-z_j)/c) \quad \text{with the}$$

function $f_+$ given by:

$$f_+(\omega, \omega_0, a) = \frac{1}{\omega_0}\left(-\cos(\omega_0 a)Ci((\omega+\omega_0)a) + Ci(\omega a) - \sin(\omega_0 a)Si((\omega+\omega_0)a)\right)$$
$$+ \frac{i}{\omega_0}\left(\sin(\omega_0 a)Ci((\omega+\omega_0)a) + Si(\omega a) - \cos(\omega_0 a)Si((\omega+\omega_0)a)\right) \quad \text{(A3-6)}$$

Moreover, for $|z - z_j| \gg c/\omega_1$ (and thus $|z - z_j| \gg c/\omega_2$), we have:

$$\wp \int B(\omega_k) \frac{e^{-i\frac{\omega}{c}(z-z_j)}}{\omega_k(\omega_k-\omega_0)} d\omega_k \simeq i\frac{2}{\omega_0} e^{-i\frac{\omega_0}{c}(z-z_j)} Si(\Delta_0(z-z_j)/2c) \quad \text{(A3-7)}$$

With $\Delta_0 = \omega_2 - \omega_1$. Since we have $\Delta_0 \gg \Gamma, \Delta \gg c/|z-z_j|$ (and so $\Delta_0|z-l|/c \gg 1$), we obtain $Si(\Delta_0(z-z_j)/2c) \simeq -\frac{\pi}{2}$. It finally results from (A3-5) the following relation:

$$A_+(t,z) = A_{refl}(t+z/c) \quad \text{(A3-8)}$$

With:

$$A_{refl}(t+z/c) = -i\frac{g_k \varepsilon_k}{c} \frac{2\pi}{\omega_0} \sum_{j=1}^{2} e^{i\frac{\omega_0}{c}z_j} \beta_j(t+(z-z_j)/c) \quad \text{(A3-9)}$$

The field then exhibits a spatial-temporal dependence in $t + z/c$ and can be identified with the reflected field (that necessarily propagates in this way) and is proportional to the population amplitudes of excited states.

For $z > 0$, the same demonstration can be established for the radiated field but with difference that no incident field comes from $z > l$.

# Figure captions

**Figure 1:** (a) Configuration of the atoms and the initial photon wavepacket in the waveguide. The dimension of the waveguide transverse section is $d$. The atoms are in the ground level and separated by a distance $l$. The resonant wavelength is $\lambda_0$.

(b) Atoms + field states involved in the interaction process with RWA or non-RWA nature of the coupling.

**Figure 2:** Quantum Paths leading to the modification of $\beta_1$, the excited state amplitude of atom1. Paths are associated with (a) absorption of the initial photon, (b) relaxation of atom 1 with emission of a photon in forward/backward directions, (c) relaxation of atom 2 with emission of a photon in the backward ($M_1$ amplitude) or forward ($M_3$ amplitude) directions and that interacts further with atom 1 (RWA terms), (d) excitation of atom 1 with emission of photon in the backward ($M_2$ amplitude) or forward ($M_4$ amplitude) directions and that interacts further with atom 2 (non-RWA terms). Similar photon diagrams exist for the modification of $\beta_2$.

**Figure 3**: Pulse-area theorem.

Temporal behavior of the transmitted field with (a) real part, (b) imaginary part. We represent the curves for several ratios $\Gamma/\Delta$. The inter-atomic distance is such as $k_0 l = \dfrac{\pi}{4}$. The temporal area vanishes in all cases. In (c) is represented the corresponding spectra and in dash- dotted line the incident spectrum. The resonance frequency is not transmitted.

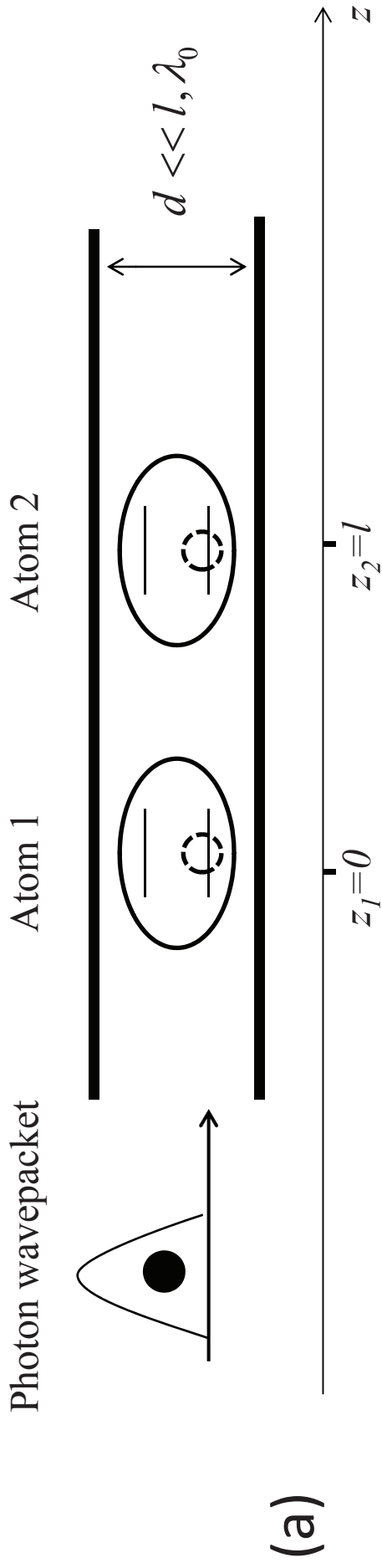
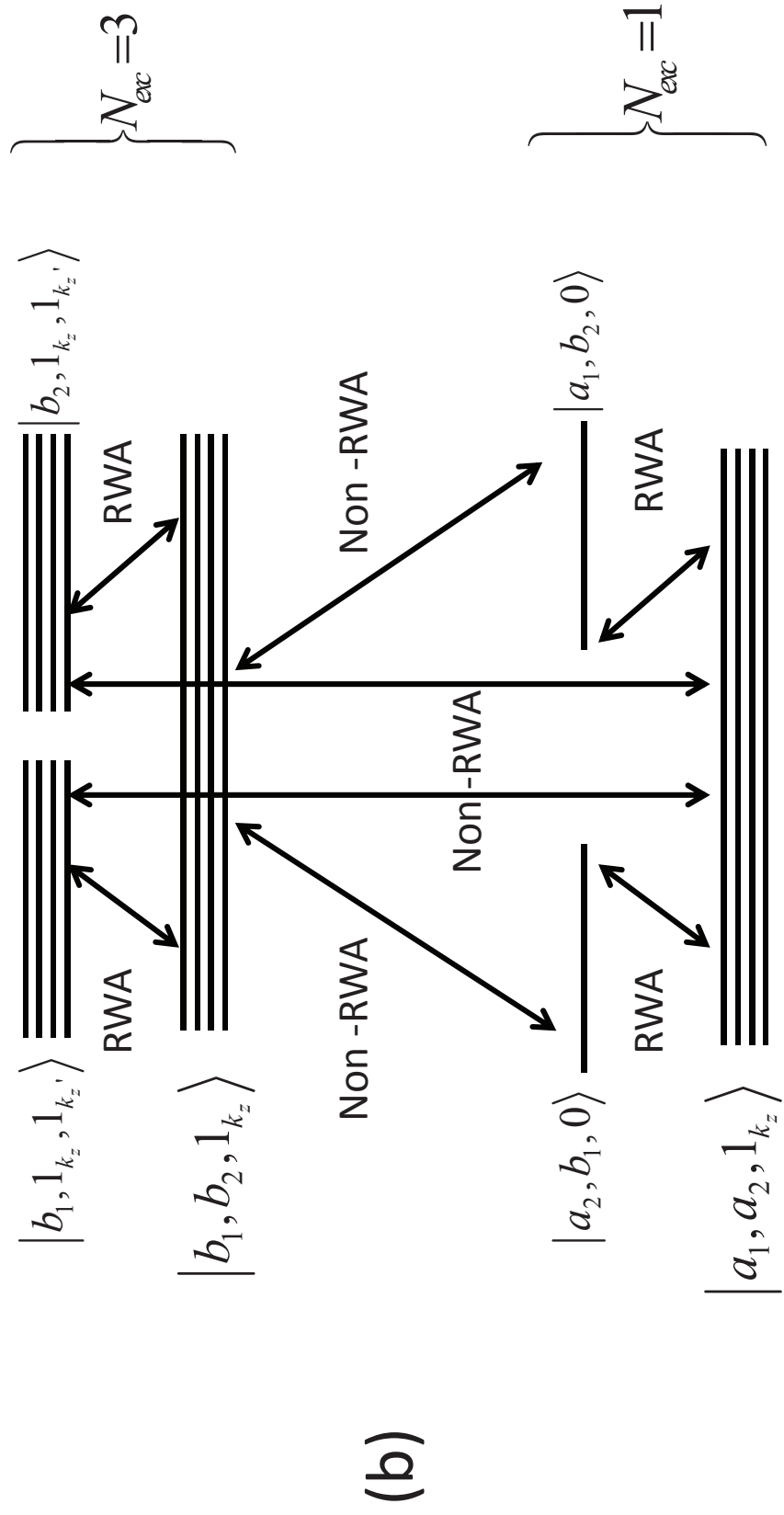

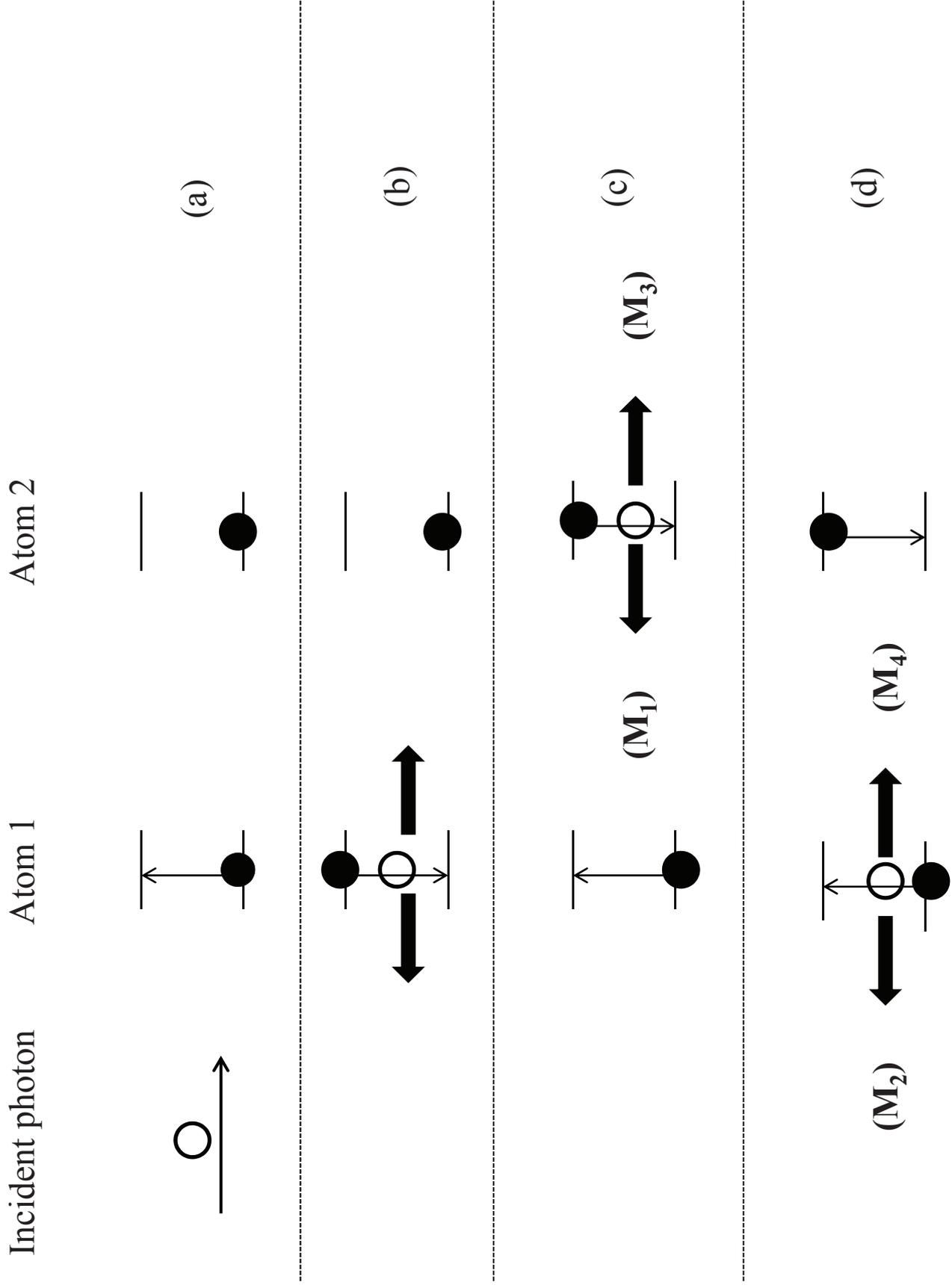

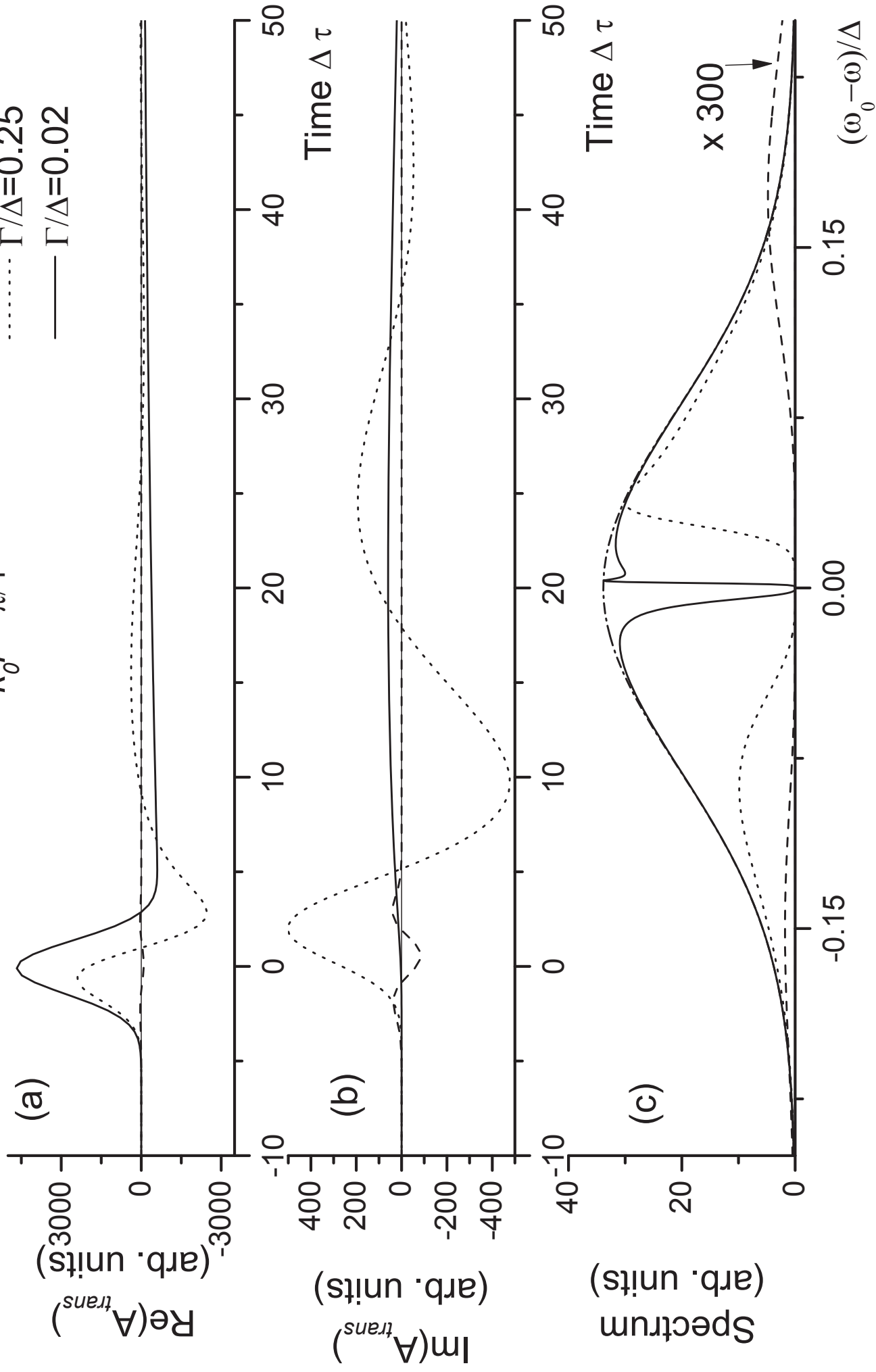